\documentclass[5p,times]{elsarticle}
\date{}

\usepackage{graphicx}
\usepackage{amsmath}
\usepackage{amssymb}
\usepackage{subcaption}
\usepackage{csquotes}
\usepackage{algorithm}
\usepackage{algpseudocode}
\usepackage[english]{babel}
\usepackage[hidelinks]{hyperref}
\usepackage{url}

\biboptions{sort&compress}

\usepackage{float}
\usepackage{stfloats}

\journal{arXiv}

\begin{document}

\begin{frontmatter}



\title{Geant4: a Game Changer in High Energy Physics and Related Applicative Fields }


\author[cern]{Tullio Basaglia\fnref{fn1}}
\author[ieee]{Zane W. Bell}
\author[dibris,infnge]{Daniele D'Agostino\corref{cor1}}
\author[ieee]{Paul V. Dressendorfer}
\author[cern]{\\Simone Giani}
\author[infnge]{Maria Grazia Pia}
\author[infnge]{Paolo Saracco}

\affiliation[cern]{organization={European Organization for Nuclear Research (CERN)},
           city={Geneva},
           addressline={CH-1211 Geneva 23}, 
            country={Switzerland}}

\affiliation[ieee]{organization={IEEE},
            addressline={445 Hoes Lane}, 
            city={Piscataway, NJ 08854},
            country={USA}}

\affiliation[dibris]{organization={Department of Informatics, Bioengineering, Robotics and Systems Engineering (DIBRIS), Università degli Studi di Genova},
            addressline={Via Dodecaneso 35}, 
            city={Genova},
            country={Italy}}

\affiliation[infnge]{organization={Istituto Nazionale di Fisica Nucleare (INFN), Sezione di Genova},
            addressline={Via Dodecaneso 33}, 
            city={Genova },
            country={Italy}}
          
\cortext[cor1]{Corresponding author, daniele.dagostino@unige.it}
\fntext[fn1]{Retired.}

\begin{abstract}
Geant4 is an object-oriented toolkit for the simulation of the passage of
particles through matter. 
Its development was initially motivated by the requirements of physics
experiments at high energy hadron colliders under construction in the last
decade of the 20$^{th}$ century.
Since its release in 1998, it has been exploited in many different applicative
fields, including space science, nuclear physics, medical physics and archaeology.
Its valuable support to scientific discovery is demonstrated by more than
16000 citations received in the past 25 years,
including notable citations for main discoveries in different fields.
This accomplishment shows that well designed software plays a key role in enabling scientific advancement.   
In this paper we discuss the key principles and the innovative decisions at the basis of Geant4,
which made it a game changer in high energy physics and related
fields, and outline some considerations regarding future directions.
\end{abstract}

%


\end{frontmatter}



\section{Introduction}
\label{sec:introduction}

Geant4 \cite{g4nim, g4tns, g4nim2} is an open source, object-oriented software toolkit for the
simulation of the interactions of particles with matter.
Its development was driven by the requirements of high energy physics 
experiments at large scale particle accelerators under construction in the last
decade of the 20$^{th}$ century.
Twenty-five years since its first public release, it is still actively
maintained and widely used in a variety of domains for scientific research, for
engineering and industrial tasks, for medical applications and for
investigations of cultural heritage.

The success of Geant4 at enabling scientific discoveries and supporting
technological advancements is demonstrated by the large number of citations of
its main reference \cite{g4nim} and of patents based on it.
These achievements are rooted in the software engineering grounds that shaped
Geant4 conception, design and development.

This paper summarizes Geant4 main characteristics, 
documents Geant4's contribution to multidisciplinary scientific results over the
past 25 years, reviews the role of software engineering at supporting them and
discusses some future perspectives in the field.
The focus is on showing that the grounds of Geant4's success and
long life without any major structural update have been careful
design, based on the exploitation of modern software engineering techniques, the
adoption of  the object-oriented methodology, and an early decision to use the
C++ language, which in 1994, at the time when Geant4 development started,
represented a brave decision.

Thanks to these choices, the software represented a game-changing factor that
has shaped research work practice and has enabled scientific progress not only in
high energy physics, where it contributed to the discovery of the Higgs boson
\cite{higgs_atlas, higgs_cms}, but also in several other research fields.

The paper is structured as follows. 
Section \ref{sec:geant4} presents Geant4 architecture and basic features.
Section \ref{sec:montecarlo} reviews the situation of software for Monte Carlo particle
transport in 1994 and today, followed by an overview of 
discoveries and scientific advancements enabled by Geant4 in Section
\ref{sec:impact}.
A discussion of key design aspects is elaborated in Section \ref{sec:sweng},
followed by some concluding remarks and considerations on future perspectives.

\section{The Geant4 Simulation Toolkit}
\label{sec:geant4}

\subsection{Geant4 Development}
\label{sec:development}

Geant4 was developed by the RD44 project \cite{rd44, rd44-95, rd44-97, rd44-98}.
Despite the similar name, it represented a radical shift
with respect to the GEANT code described in Section \ref{sec:geant3}.
RD44 gathered an international team of physicists, computer scientists and engineers
affiliated with many academic and research institutes worldwide.
It took part in the CERN (Conseil Europ\'een pour la Re\-cher\-che Nucl\'eaire)
research and development programme for the experiments \cite{virdee_2004} at the
Large Hadron Collider (LHC) \cite{price_2002}, which were expected to become
operational in the first decade of the 21$^{st}$ century.

Early considerations about developing an object-oriented Monte Carlo system for
particle transport date back to 1993 \cite{takaiwa_1993, giani_1993}.
They evolved into the formal proposal \cite{p58} of Geant4 development, which was
approved by the CERN Detector Research and Development Committee (DRDC) as the RD44 pro\-ject in
November 1994.
Geant4 was first released on 15 December 1998.
The RD44 members are listed  in  \ref{app:rd44}.

At the end of the RD44 research and development phase, the maintenance of the
Geant4 toolkit was formally taken over by an international team known as the
Geant4 Collaboration, which is responsible for releasing new versions of the
code and for providing user support.

\subsection{Basic Features of Geant4}
\label{sec:g4basics}

Geant4 is an object-oriented software toolkit for the simulation of the passage
of particles through matter. 
It tracks particles and models the processes through which they interact with
the environment they traverse.
Geant4 source code and the associated user documentation can be freely
downloaded \cite{g4web}.

Simulation plays a fundamental role in high energy physics experiments -- the
scientific domain that motivated the development of Geant4.
It permeates the whole experimental life-cycle: from the conceptual design of
experiments and the optimization of the characteristics of the particle
detectors they encompass, to in-depth understanding of their operation, down to
the evaluation of various physics and technological factors that contribute to
determine the scientific results of an experiment.
It is also an essential basis for the development and the refinement of 
software for data reconstruction using signals recorded by particle detectors,
and for data analysis.
Similar functions are also common to other experimental scenarios where precise
account of the effects of particle interactions with matter is needed, such as
space science missions and astrophysics, nuclear power plants, radiation therapy,
radiation protection, etc. 

\begin{figure}
\centerline{\includegraphics[angle=-90,width=6.2cm]{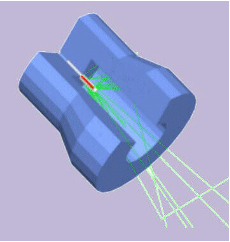}}
\caption{Visualization of a Geant4-based simulation of a radioactive source and
its enclosure for application in oncological radiation therapy: the geometrical
model reproduces the container of the radioactive source (in red) and the
applicator (in blue) used in clinical practice; primary and secondary particles
originating from the radioactive source are shown in green. }
\label{fig:brachy}
\end{figure}

\begin{figure}
\centerline{\includegraphics[angle=0,width=8cm]{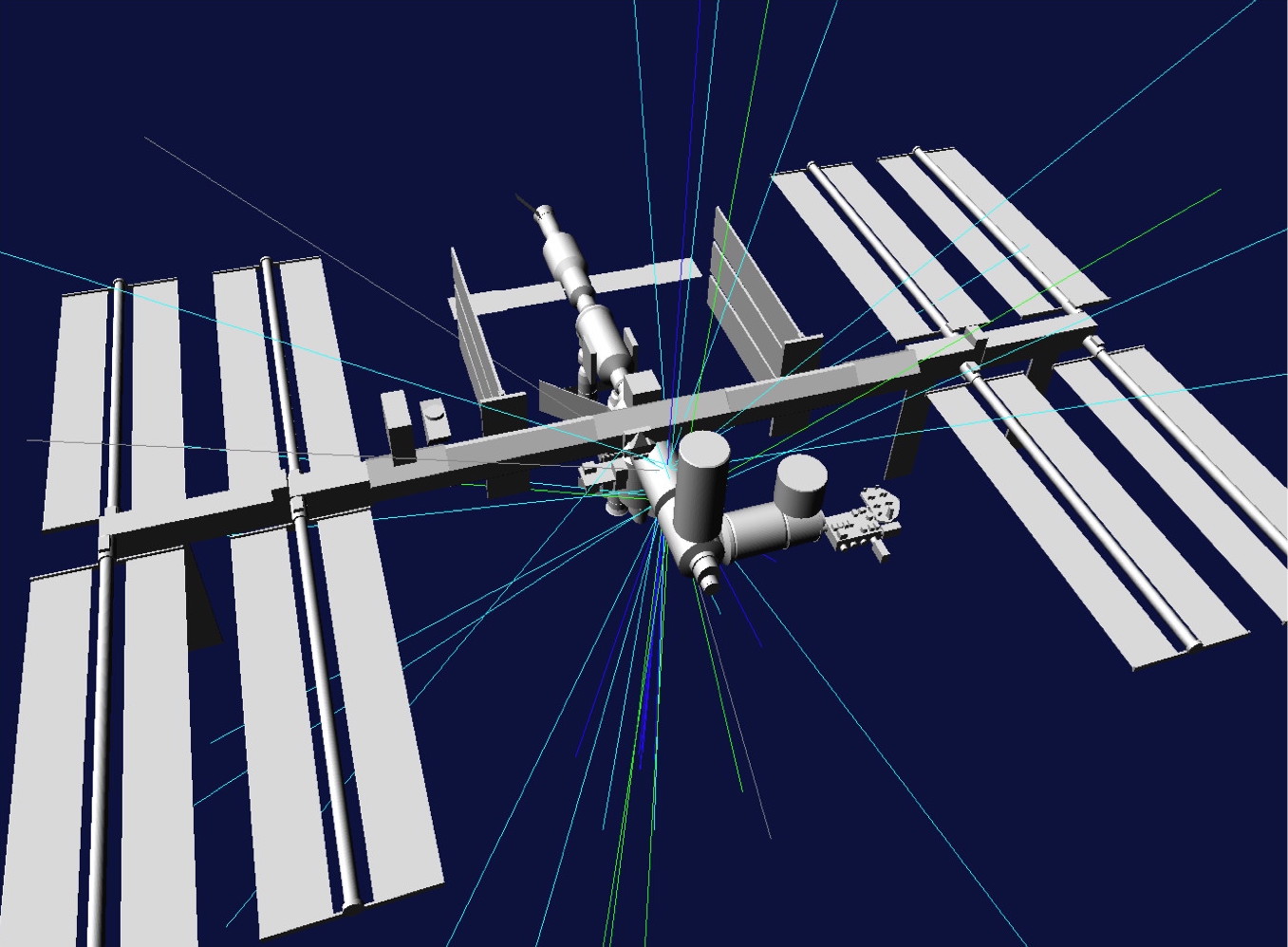}}
\caption{The International Space Station modelled with Geant4.}
\label{fig:issg4}
\end{figure}

Geant4 includes tools for all the functionality typically nee\-ded for
the simulation of these experimental scenarios: modeling the geometry and 
materials of the experimental setup, retrieving the properties of elementary and
composite particles, modeling the processes particles undergo when interacting
with matter, managing the passage of particles in the experimental setup
(articulated over steps, tracks and events)
and recording its effects,
controlling the execution of the simulation, visualizing the experimental setup
and particle trajectories, and providing user interface to the simulation.
Additionally, the Geant4 toolkit includes a set of auxiliary tools supporting the 
core functionality and a rich collection of application examples.
Further details can be found in the main Geant4 reference paper \cite{g4nim}.

The simulation scenarios illustrated in Figures \ref{fig:brachy} and \ref{fig:issg4},
pertaining to utterly different application fields and characterized by largely
different scales, are suggestive of the versatility of Geant4.
Figure \ref{fig:brachy} reproduces the model of an $^{192}$Ir radioactive source
used in oncological radiation therapy \cite{foppiano_2003}. The size of
brachytherapy applicators, such as the one depicted in the figure, is of the
order of a few centimeters.
Figure \ref{fig:issg4} shows the Geant4-based model of the International Space
Station \cite{ersmark_2007} exposed to the space radiation environment.


\subsection{The Architecture of Geant4}
\label{sec:g4architecture}

Geant4 is a toolkit: a set of consistent software tools, out of which
the user picks those deemed relevant to the experimental problem to be investigated.
As a toolkit, Geant4 cannot be ``run'' by a user: the user must develop an
application, specific to the experimental scenario one intends to simulate,
which uses the ``tools'' available in the toolkit.
This is a 
a distinctive characteristic of Geant4, which distinguishes it from other Monte
Carlo particle transport codes, such as those briefly described in Section
\ref{sec:montecarlo}.

\begin{figure*}
\centerline{\includegraphics[angle=0,width=16cm]{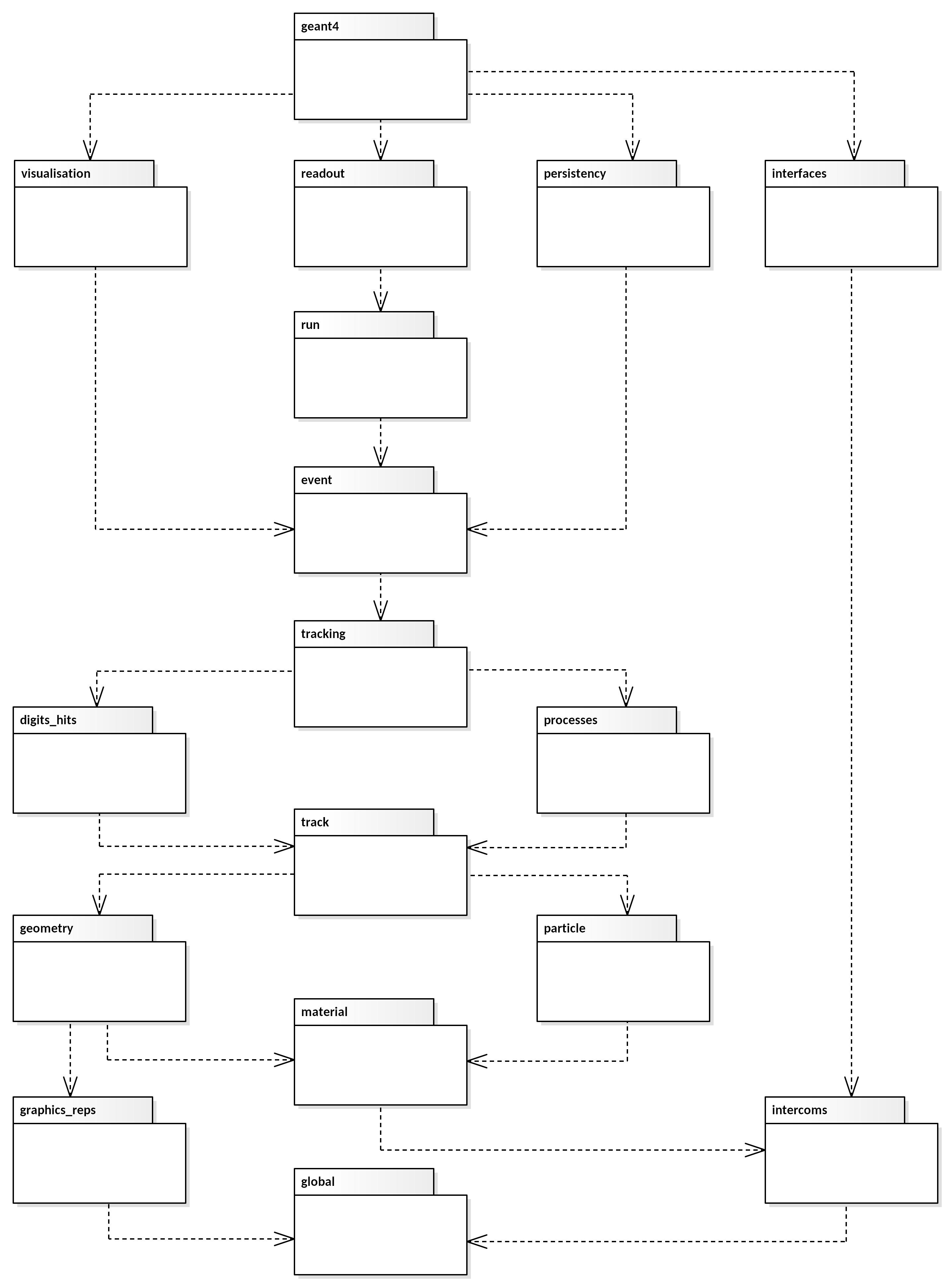}}
\caption{UML package diagram of Geant4, showing the top level packages. Further details and UML diagrams 
of Geant4 software design can be found in \cite{g4nim}. }
\label{fig:g4uml}
\end{figure*}

The foundation of Geant4 lies on a rigorous problem domain analysis and related
object-oriented design, conceived and progressively refined during the course of the
RD44 project, which identified well-defined interfaces and relations between
components, making the toolkit easily and consistently extensible.
The Geant4 architecture resulting from this process is illustrated
in the UML (Unified Modeling Language) package diagram of Figure \ref{fig:g4uml}.

Packages at the bottom of the diagram in Figure \ref{fig:g4uml} provide fundamental services
to the whole toolkit:
\textit{global} deals with basic features (the system of units, physical
constants, mathematical methods and random numbers); \textit{material} and
\textit{geometry} deal with modeling the experimental setup; \textit{particle}
encodes particle properties \cite{pdg_2022}; \textit{graphic\_reps} and \textit{intercoms}
enclose basic means for graphics and interactions with Geant4 kernel.
Packages above them deal with the core of particle transport: \textit{track}
hosts the key players of transport -- classes responsible for tracks and steps; \textit{processes}
deals with particle interactions in the course of transport; \textit{digits\_hits}
deals with the generation of the responses to physics processes in the
experimental setup; \textit{tracking} manages the evolution of particle
transport and of its effects in sensitive volumes of the experimental layout.
A notable feature of Geant4 architecture is that tracking deals with all
processes blindly, i.e. through a single abstract base class: this distinctive
feature is the key to the extensibility of Geant4 physics capabilities.
The \textit{run} and \textit{event} packages deal with the control of the simulation:
in the context of Geant4, an ``event'' is the minimal unit of simulation,
related to the evolution from the creation of primary particles fed into the
simulation to the resulting observables in the experimental setup; a ``run''
manages collections of events characterized by a common experimental scenario.
The  \textit{interfaces}, \textit{visualisation} and \textit{persistency} packages
depend on previously mentioned packages and let the user connect to external facilities 
through abstract interfaces (i.e. without any external dependencies) to
steer the simulation, to produce visual representations and to store persistent products, respectively.
A detailed description of Geant4 architecture can be found in the main associated
reference \cite{g4nim}.

It is worthwhile to note that the architecture depicted in Figure
\ref{fig:g4uml} has remained unchanged through the 25 years' lifetime of Geant4,
while the source code evolved from 
about 283000 lines (excluding comments)  in the first version
released in December 1998 to  
more than 872000 lines of the most recent version, 11.2, released in December 2023.
The resilience of Geant4 architecture to the extensive evolution of the software
and of its applications is a 
demonstration of the depth of the
problem domain analysis and the vision that characterized the RD44 project.

\begin{figure*}
\centerline{\includegraphics[angle=90,width=0.85\textwidth]{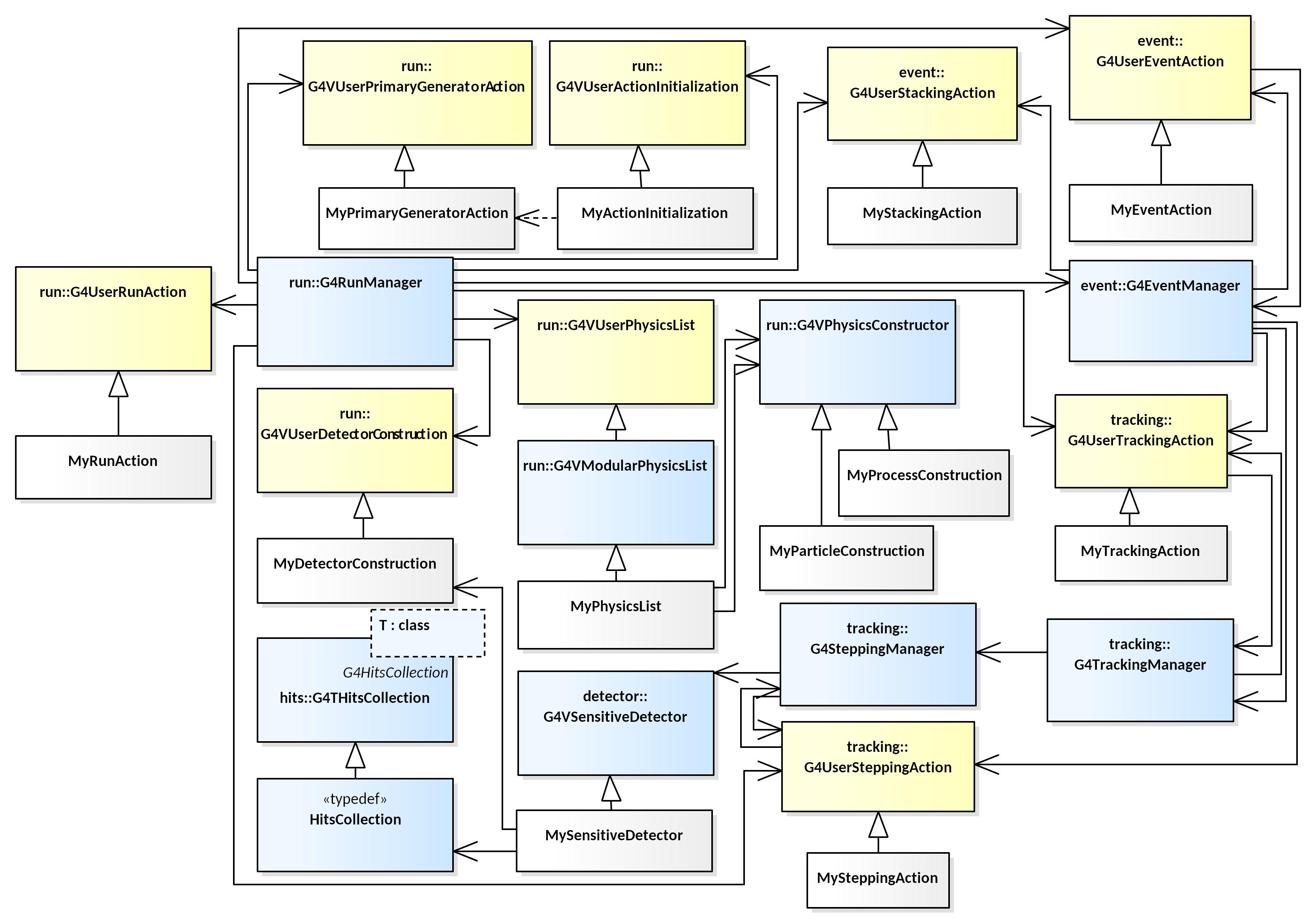}}
\caption{An example of UML diagram of a simple Geant4-based simulation application: user classes are in grey, 
relevant simulation management classes in Geant4 kernel appear in blue and base classes in Geant4 kernel
for user actions and initializations appear in yellow.}
\label{fig:g4app}
\end{figure*}

The UML class diagram in Figure \ref{fig:g4app} illustrates the conceptual
organization of a Geant4-based user application.
The user instantiates \textit{G4RunManager}, a Geant4 kernel class responsible
for orchestrating particle transport in the user-defined experimental scenario,
and interacts with the Geant4 kernel in the course of the simulation through 
a set of initialization and action classes, derived from base classes present in 
Geant4 kernel.
These classes, shown in yellow in Figure \ref{fig:g4app}, allow the user to
customize the simulation.
They are responsible for the creation of the primary
particles to be fed into the simulation, for modeling the experimental setup,
for selecting the physics processes and models relevant to the experimental
problem to be investigated, and for the interaction with the transport kernel at
various stages (run, event, tracking, stepping etc.).
An overview of the workflow of using Geant4 in a simple simulation can be found
in Geant4 user documentation \cite{geant4_appdevmanual112}.

While the class diagram in Figure \ref{fig:g4app} reflects the conceptual
configuration of any Geant4-based simulation, the simulation of large-scale
experiments using Geant4 is articulated through complex software designs and
implementations,  
such as, for instance, those described in \cite{atlassimulation,
banerjee_2012, clemencic_2011}.

The most critical feature of the simulation of an experimental scenario is the
selection of the relevant physics processes, cross sections and final
state generation models among the many options available in the Geant4 toolkit.
This operation is supported by validation tests of Geant4 physics 
modeling elements (e.g. \cite{wellisch_2003, tns_photoel, tns_photoel2, tns_beb}),
complemented by experiment-specific comparisons of the simulation of complex
observables (e.g. \cite{natochii_2021, chefdeville_2019, morgunov_2021})
to corresponding experimental measurements.
The validation process allows users to optimize their simulation
configuration and to evaluate its contribution to the errors associated with
their experimental measurements \cite{saracco2017propagation}.

\subsection{Computational Performance}
\label{sec:performance}

The computational performance of a Geant4-based simulation intertwines the
intrinsic performance of Geant4 elements, the way they are used, and the
performance of the user application.
Therefore, the performance of a Geant4-based simulation can only be quantified
over the specific configuration defined in the application, together with
possible inefficiencies in the application code \cite{kourlitis2023optimizing}.

An exhaustive discussion of Geant4 computational performance exceeds the scope
of this paper.
In this brief overview, we highlight some features that have played a
significant role in the production of scientific results in
computationally-intensive simulation applications.

Computational performance has been a concern since the early days of Geant4 
development. 
It was addressed in the RD44 project through a combination of conceptual
innovations in the software design and in the implemented algorithms, along with
attention to the quality of the software through peer reviews and frequent
benchmarks to monitor and optimize the performance of the code.

A common prejudice in the  
1990s was the unsuitability
of C++ as a programming language for computationally-intensive scientific use.
Early benchmarks \cite{rd44-95} demonstrated that, thanks to well designed and
carefully implemented software, Geant4 achieved much better computational
performance than GEANT 3.21, which had been purposely optimized for these assessments, 
in simulation scenarios specifically devised for performance evaluations.

A major innovation in terms of computational performance was the introduction of
the concept of ``smart voxels'' in Geant4 geometry \cite{kent_smart_1995}, which
significantly improves the performance of tracking particles in the experimental
setup, i.e. a most critical task in the simulation of any experimental scenario.
Geant4 ``smart voxels'' extend and enhance the established concept that a
hierarchically structured geometry contributes to computational performance of
tracking time through the reduction of candidate volumes for intersections.
A relevant feature of their implementation is the low computational and memory cost.
The optimization realized by this algorithm even reduces the need for revising 
poorly structured geometries coded by inexperienced users in their applications.
Further actions to optimize the performance of tracking, both in Geant4 software
design and implementation, are documented in \cite{rd44-97}.

Another innovative algorithm \cite{open_99299} exploits Geant4's  
unique method of dealing with thresholds for secondary particle production in
terms of particle range, rather than on tracking cuts or energy cuts as in
other Monte Carlo transport codes. 
The adoption of secondary particle production in terms of range thresholds
simultaneously addresses computational performance and physical accuracy in the
occurrence of electromagnetic processes (ionisation and Bremsstrahlung) affected
by infrared divergence. 
It lets processes produce secondary electrons
($\delta$-rays) that would reach geometrical boundaries, thus having a chance to
produce observable effects, while it avoids the waste of computational resources
for producing and tracking low energy secondary electrons that would not escape
from the current volume by locally depositing their energy.
As shown in \cite{open_99299}, it
significantly optimizes the tracking performance for ionizing particles without
sacrificing physics accuracy. 
Further details about performance optimization resting on Geant4 design and
algorithms can be found in \cite{rd44-95}, \cite{rd44-97} and \cite{g4nim}.

A key factor improving the performance of Geant4 is represented by the adoption
of the multithreading paradigm, introduced in version 10 in 2013 \cite{dong_2010, dong_2012, g4nim2}
in response to the emergence of multi-core and many-core processors \cite{7581868}.
The foundation of this evolution relies on the intrinsic independence of events
in Geant4, embodied by the event management originally designed in RD44, which
made Geant4 naturally parallelizable.
Each thread is responsible for simulating one or more events, thus implementing
event-level parallelism; memory savings are obtained by sharing data that remain
constant throughout the simulation, such as the geometry model and data used by
physics processes.

This solution addresses current experimental requirements; further
investigations are in progress in view of the evolution of the requirements of
the experiments at the High Luminosity LHC \cite{hllhc_2020}, whose operation is
currently foreseen in the years 2030-2041, and of the evolution of parallel
architectures in the hardware industry.
Research on this computational topic involves both experimental teams and
members of the Geant4 collaboration, and concerns porting the code to GPU
(Graphics Processing Unit) devices \cite{amadio2023offloading} as well as the investigation of other
candidate solutions.

\section{Related Works}
\label{sec:montecarlo}

\subsection{GEANT}
\label{sec:geant3}

Geant4 superseded the GEANT \cite{geant1, geant321} 
simulation tool previously produced and distributed by CERN.

The GEANT (GEometry ANd Tracking) program was developed as a detector
description and simulation tool for particle physics experiments.
It was written in Fortran; its development represented at least 50 man-years'
work, spread over more than 15 years.
The early developments of GEANT did not materialize into proper releases
of the code; the first practically usable release was GEANT version 3 in 1984,
which evolved through a series of minor versions.
The last version, GEANT 3.21, consisting of approximately 63000 lines of code,
was released in early 1994; 
it was maintained by CERN without any further development until 2003.
 
The main functionality of GEANT consisted of tracking particles through a
geometrical data structure.
While its geometry and tracking capabilities were 
refined, GEANT was
limited in modeling physics processes.
It included functionality to handle basic electromagnetic interactions of
particles with matter, but it relied on external packages (GEISHA
\cite{fesefeldt_1985}, FLUKA \cite{ranft_1980} and CALOR \cite{calor_1977}) for
the simulation of ha\-dro\-nic interactions.
The dependency on critical physics code, owned and controlled by external
sources, de facto prevented the improvement and the extension of GEANT
capabilities in the hadronic physics domain as needed by the evolving 
requirements of the experiments at CERN and elsewhere.
The electromagnetic physics of GEANT lacked adequate modeling capabilities for
precision quantification of the effects of particle interactions, especially at
very low ($<10$ keV) and very high ($\gtrsim100$ GeV) energies, which are relevant to application fields such
as astroparticle physics, medical physics and space science.

A serious drawback of procedural programming in GEANT 
(common to other simulation codes in Fortran) was the difficulty in
extending its functionality in response to 
requirements that commonly arise in physics research: for instance, nearly 60
subroutines had to be modified, when one wished to implement an additional
geometrical shape needed to model an experimental setup.
One had to face a similar complication when introducing a new particle type to
be tracked.

The success of GEANT in experimental particle physics was largely due to allowing
users to plug in routines for handling ex\-pe\-ri\-ment-dependent code, such as the
geometrical description and the digitisation of the signal of detectors, into an
infrastructure of experiment-independent code.
Nevertheless, the functionality of GEANT was inadequate for 
the long-term research programs of the experiments in
the TeV scale that were being designed in the 1990's 
 and expected to be operational after 2000. 
The use of GEANT in research areas other than particle physics, such as space
science \cite{malaguti_1991} and medical physics \cite{michel_1991}, was
marginal in the mid 1990's.

\subsection{Overview of Monte Carlo Particle Transport Codes}
\label{sec:mccodes}

Many software systems have been developed for the simulation of particle
interactions with matter since the 1940's, when the Monte Carlo method was
devised and first applied to physics calculations on early computers
\cite{metropolis_1949, vonneumann_1951, eniac_2014}.

\begin{figure} [htbp]
\centerline{\includegraphics[angle=0,width=0.99\columnwidth]{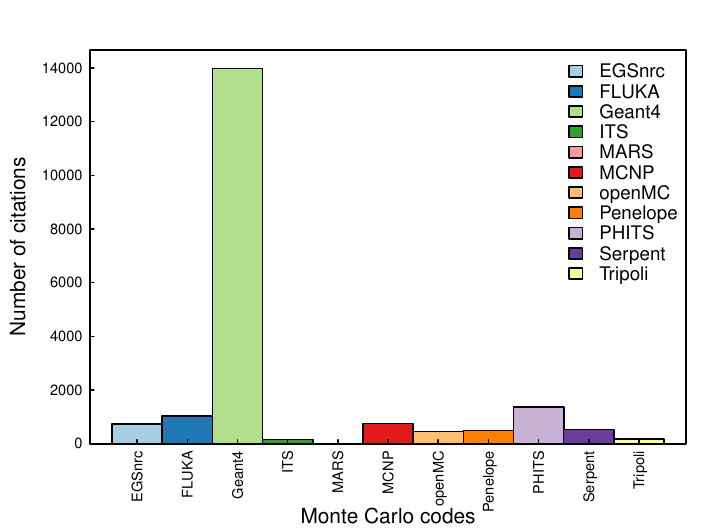}}
\caption{Citations of Monte Carlo codes commonly used in particle, astroparticle,
nuclear and medical physics, in space science and related research areas, collected over
the years 2003-2023. }
\label{fig:mccitations}
\end{figure}

Several Monte Carlo particle transport codes are currently utilized in particle,
astroparticle, nuclear and medical physics, in space science and related fields
along with Geant4.
Well-known codes in these domains are
EGS5 \cite{egs5},
EGSnrc \cite{egsnrcJournal}, 
FLUKA \cite{fluka11}, 
ITS \cite{its3},
MARS \cite{mars},
MCNP 6 \cite{mcnp6},
OpenMC \cite{openmc1},
PENELOPE \cite{penelope1995},
PHITS \cite{phits2002},
Serpent \cite{serpent2013} and
TRIPOLI-4\textsuperscript{\textregistered} \cite{tripoli4})
Some of them have a fairly wide scope of applicability (e.g., MCNP), while
others are restricted to specific physics scenarios (e.g. handling only electron
and photon interactions) or are targeted to specific experimental application
areas (e.g. simulations related to nuclear reactors).
Some have been well established in the field for decades (e.g. MCNP, ITS and
EGS), while others are relative newcomers (e.g. PHITS, Serpent and OpenMC).
Most of them are still written in Fortran; some include recently developed parts 
in C++ along with long-standing Fortran code.

A detailed description of the features of these codes and a discussion of their
relative merits is well  beyond the scope of this paper;
nevertheless, some considerations are helpful to best appraise the role played 
by Geant4 in enabling the production of scientific results.

Geant4 stands out among Monte Carlo particle transport codes with respect to the
number of citations received by the associated reference papers, as is shown in 
Figure \ref{fig:mccitations}.
The citation data used in the plot derive from  
the Web of Science\textsuperscript{\texttrademark} \cite{wos}
and are limited
to publications in scholarly journals appearing between 2003 and 2023, inclusive.
It should be noted that only a few of the associated references were published
in 2003 or earlier; more recently developed codes have necessarily accumulated
citations over a reduced portion of the time scale of the plot,
nevertheless, the variable extent of collection of citations does not
substantially affect the qualitative appraisal of Figure~\ref{fig:mccitations}.

It is also worthwhile to note that one of the previously mentioned Monte Carlo
codes, EGS5, does not appear in the plot due to lacking a dedicated journal
publication we could use to extract scientometric data 
(note that \cite{egs5} is not a peer-reviewed document). 
Indeed, user documentation manuals have been for a long time the only writings
associated with Monte Carlo transport codes; even codes with a long historical
background began publishing dedicated reference articles only relatively recently.
This habit reflects the unfounded perception of Monte Carlo transport
codes, and of scientific software in general, as mere service tools devoid of
scientific relevance in some physics and engineering environments, which 
was quantitatively assessed in \cite{swpub}.
This attitude is responsible for the frequent omission of proper citation of the
existing reference papers of Monte Carlo codes, including Geant4, which was
highlighted in \cite{basaglia_2017}.

\section{The Impact of  Geant4 in the Scientific Community}
\label{sec:impact}

The Geant4 main reference paper has collected more than 16000 citations by the end
of 2023 
in the Web of Science\textsuperscript{\texttrademark}, 
here considered since 1990.
It is the most cited publication in the categories of Particle and Fields
Physics, Nuclear Physics, Nuclear Science and Technology, Instruments and
Instrumentation, 
and Astronomy and Astrophysics.
These categories collectively include more than 2~million papers
published in scholarly journals between 1990 and 2023.
These data concisely assess the impact of Geant4 on the production of
scientific results.

\begin{figure}[hbt]
\includegraphics[width=\columnwidth,trim=2 0 7 20,clip]{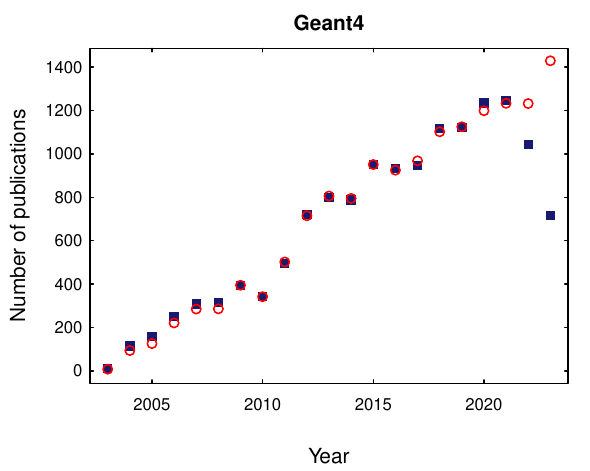}
\caption{Number of scientific publications citing Geant4 main reference \cite{g4nim} as a function of year:
data retrieved from the Web of Science (blue squares) and from Scopus (red circles).}
\label{fig:g4papers}
\end{figure}

Figure \ref{fig:g4papers} shows the time profile of the citations of
Geant4 main reference \cite{g4nim};
it reports data retrieved from the Web of
Science\textsuperscript{\texttrademark} and from
Scopus\textsuperscript{\textregistered} \cite{scopus},
concerning journal publications.
The two data sets look generally consistent, but some discrepancies
are qualitatively visible in the number of citations of the last two years.
This apparent anomaly could be related to ongoing updates and corrections
of the database underlying the Web of Science; nevertheless, being limited
to a small fraction of the data, it does not substantially affect the 
scientometric results reported in this paper.
Mann-Kendall \cite{mann_1945, kendall_1948} trend tests performed over both
sets of citation data, including the whole range of years from 2003 to 2023, reject
the null hypothesis of no trend in the data distributions with 0.01 significance
in favour of the alternative hypothesis of increasing trend, thus supporting the
irrelevance of the two last points in determining the general trend of the data.

\begin{figure}[hbt]
\includegraphics[width=0.97\columnwidth,trim=2 0 7 20,clip]{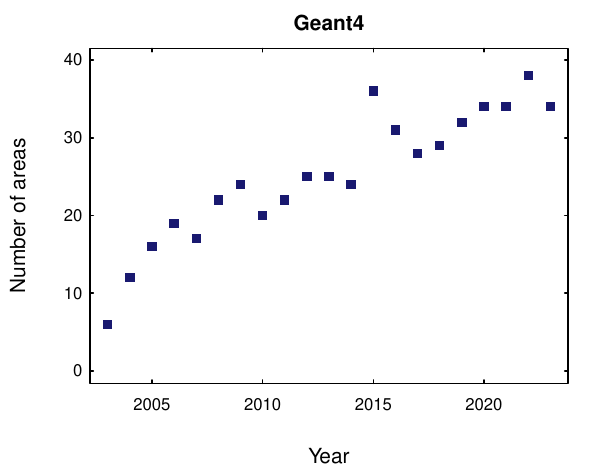}
\caption{Research areas of the publications citing Geant4 main reference \cite{g4nim} as a function of year.}
\label{fig:g4area}
\end{figure}

Although originally motivated by the requirements of high energy physics experiments, Geant4
encompasses functionality relevant to other research domains.
The literature documenting the use of Geant4 is characterized by wide and 
growing multisciplinarity: one can observe it in Figure \ref{fig:g4area}, which
shows the number of research areas (as defined in the Web of Science)
represented in the papers that cite Geant4 
main reference \cite{g4nim}, as reported by the Web of Science.
The increasing multidisciplinarity in the scientific results enabled by Geant4
is objectively confirmed by statistical inference: the Mann-Kendall test
\cite{mann_1945, kendall_1948} rejects the null hypothesis of no trend in the
data distribution of Figure \ref{fig:g4area} with 0.01 significance in favour of the alternative
hypothesis of increasing trend.

\begin{figure}[hbt]
\includegraphics[width=0.97\columnwidth,trim=2 0 7 20,clip]{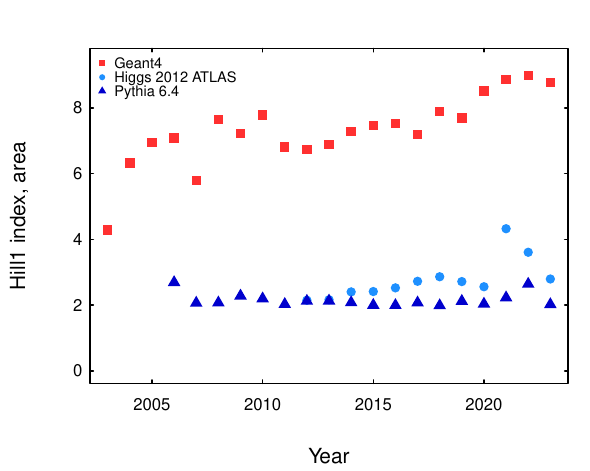}
\caption{Diversity, as a function of year, of the research areas of origin of
the citations of Geant4 main reference \cite{g4nim} (red squares) and of
representative high energy physics publications: the report of the discovery of
the Higgs boson \cite{higgs_atlas, higgs_cms}, related to the 2013 Nobel
Prize, and the reference paper of the PYTHIA Monte Carlo event generator
\cite{pythia_2006} (blue circles and triangles, respectively). 
Since the two papers \cite{higgs_atlas, higgs_cms}
were published simultaneously by the ATLAS and the CMS experiments and are
usually jointly cited, the plot only displays the diversity associated with
\cite{higgs_atlas} for better clarity.}
\label{fig:hill1area}
\end{figure}

The multidisciplinary character of the research publications that use Geant4 is
further corroborated by the analysis of their diversity.
Biodiversity is a concept pertinent to quantitative ecology, where it measures the
richness and the complexity of a community, taking into account the number of
species it hosts and their abundance. 
Its mathematical formulations \cite{roswell_2021} can be applied to the analysis
of scholarly publications.

Figure \ref{fig:hill1area} reports the time profile of a well established
diversity measure, calculated over scientometric data retrieved from the Web of
Science -- the Hill index of order 1 \cite{hill_1973}, which is related to the definition
of Shannon entropy \cite{shannon_1948} in information theory.
The plot shows the diversity of research areas of the citations of Geant4
main reference \cite{g4nim},
measured each year over the papers published in that year,
and of two significantly representative publications
originating from the same domain as Geant4, high energy physics: the
observation of the Higgs boson at the LHC, related to the 2013 Nobel Prize in
Physics \cite{higgs_atlas, higgs_cms}, and the reference paper of the PYTHIA
\cite{pythia_2006} Monte Carlo event generator, a highly cited software system
used in particle physics.
It is evident in Figure \ref{fig:hill1area} that the scientific production
enabled by Geant4 exhibits much larger diversity than either of those two papers.

It is practically impossible to document a complete assessment of the scientific
research that Geant4 has enabled and of the results that it has contributed to
achieve over the past 25 years since its first release.
We just single out a few remarkable examples that highlight the
multidisciplinary usage of Geant4, in addition to the previously
mentioned use of Geant4 in relation to the first observation
of the Higgs boson by the ATLAS and CMS experiments \cite{higgs_atlas,
higgs_cms}.

Geant4 played a crucial role in the events that affected the operation of two
similar space missions for X-ray astronomy, Chandra and XMM, launched by NASA
and ESA (European Space Agency), respectively, in 1999, shortly after the first
release of Geant4.
Chandra experienced unexpected degradation of the majority of the CCDs
(Charge-Coupled Devices) in the ACIS (Advanced CCD Imaging Spectrometer)
instrument on board shortly after the launch, which affected the scientific
capabilities of the mission.
The Geant4-based simulation of XMM \cite{nartallo2000radiation} prior to its
launch identified the cause of the deterioration of Chandra's detection
capabilities, due to low energy ($\sim$100 keV) protons reaching the focal plane
after scattering through the mirror shells in the experimental setup.
This result allowed ESA to take proper countermeasures prior to the launch to
avoid similar damage to XMM, which preserved as-designed detection
capabilities during its operation in orbit.

Geant4 has been instrumental in enabling a major archeological discovery in
Khufu's (Cheops) pyramid \cite{morishima2017discovery} on the Giza plateau in
Egypt by means of the muon tomography technique.
A large void above the Grand Gallery of the Pyramid has been identified by
comparing the expected muon flux, determined by a Geant4-based simulation, and
experimental measurements.

Geant4 is widely used in medicine for radiation therapy, radiology, nuclear 
medicine and radiation protection research; 
articulated application tools have been developed to facilitate its use 
for medical investigations.
One of them, GATE (Geant4 Application for Tomographic Emission) has become a
de~facto standard for medical imaging simulations; its reference paper
\cite{gate_2004} has collected more than 1600 citations in scholarly
publications, which demonstrate the extensive 
contribution of Geant4 to produce research results in this field.

\begin{figure}[tbh]
\centering
\includegraphics[width=.99\columnwidth,trim=0 0 0 20,clip]{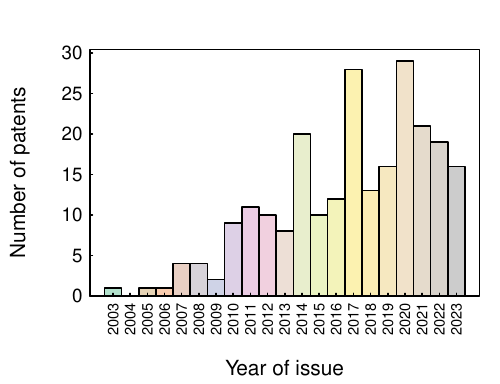} %
\caption{Patents associated with Geant4 issued by  the 
United States Patent and Trademark Office (USPTO).}
\label{fig:patentsUS} 
\end{figure}

\begin{figure}[tbh]
\centering
\includegraphics[width=.99\columnwidth,trim=0 0 0 20,clip]{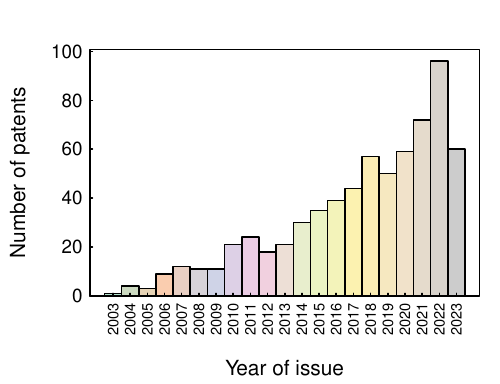} %
\caption{Patents associated with Geant4 issued by  the European Patent Office.}
\label{fig:patentsEU} 
\end{figure}

Besides academic research, Geant4 has enabled many applications of industrial relevance.
As an example, more than 200 patents related to Geant4 have been issued by the
United States Patent and Trademark Office (USPTO) between 2002 and 2023, and
more than 600 by the European Patent Office;
their distributions are illustrated in Figures \ref{fig:patentsUS} and \ref{fig:patentsEU}.
The visible growing trend in the number of Geant4-based patents issued each year
is confirmed by the result of the Mann-Kendall test \cite{mann_1945,
kendall_1948}, which rejects the null hypothesis of no trend in both data
distribution with 0.01 significance, in favour of the alternative hypothesis of
increasing trend.

\section{Software Engineering Foundations}
\label{sec:sweng}

The success of Geant4 at enabling scientific and industrial research results is 
rooted in its software engineering foundations set by the RD44 project.

RD44 exploited the most advanced software engineering techniques available at
the time of its activity (1994-1998) to accomplish a distributed
software design and development.
It introduced significant innovations in the development of Geant4 with respect
to common practice in the computational environment of high energy physics and
related physics research areas, and in the domain of Monte Carlo simulation for
particle transport.
The role of software engineering as a key strategy to produce a high-quality
scientific software system and to release it in time represented a major novelty
in these fields.


\subsection{Open Source Code}

Geant4 openness and transparency have been major concerns of the
RD44 project since its initial conception.
Open access to the source code is a necessity for physics software.
Since the
simulation software affects the outcome of the experiments that use it, 
ability to examine, understand and appraise the source code is a critical
requirement for the validation of the experimental results and, ultimately, for
their reproducibility, which is an essential feature of the scientific method.

Geant4 is freely available as open source software.
It is worthwhile to note that open access to the source code is not general
practice for Monte Carlo particle transport systems: for instance, the standalone
FLUKA \cite{fluka11} code is not publicly distributed and the MCNP \cite{mcnp6}
code is subject to  the export control laws of the United States
restricting the distribution of the source code.

Apart from its essential epistemological role in the production of scientific
results, the open source character of Geant4 code plays an important role
in allowing scientists to extend the functionality of the toolkit. 
Several users' extensions and improvements have been incorporated into the
Geant4 toolkit over the years, to the benefit of the whole user community.

\subsection{Turnarounds in Particle Physics Computing}

At the start of RD44, two paradigms had dominated the
computational environment of particle physics since the 1960's: the Fortran
programming language and the use of mainframes.

Procedural programming and Fortran (with the MORTRAN variant, a Fortran
extension used by EGS) were the rule in particle physics and in Monte Carlo
particle transport in general at the time when RD44 proposed the development of
Geant4 in 1994.
The CERN Program Library (CERNLIB) \cite{cernlib}, which included the GEANT~3
code, was for decades one of the largest and most advanced collections of
scientific software. 
As documented in \cite{cernlib_1979} and \cite{cernlib_2003}, between 1979 and
the end of its support in 2003, 80\% of its programs were written in Fortran and
the remainder in assembler code or C, usually with a Fortran version also
available.

Geant4 represented the first attempt to redesign a major package of CERN
software for an object-oriented environment.
The adoption of the object-oriented paradigm and of C++ as programming
language for Geant4 represented a breakthrough in the particle physics
environment and in Monte Carlo particle transport.

The choice of the object-oriented technology broadened the horizon and
generalized the potential for scientific research with respect to any previous
software development project in the high energy physics environment.
One could envisage dealing transparently with any experimental scenario, 
regardless of whether it was a particle detector or the human body, or propagating 
particles through any field - magnetic, electric or gravitational.
The unbounded versatility associated with the object-oriented technology
introduced an additional requirement of generalization and flexibility into
Geant4 software: the simulated physics processes must deal with an unprecedented
energy range, from approximately 100~eV to the scale of PeV, i.e. orders of
magnitude above and below the energies addressed by GEANT 3 and other Monte
Carlo particle transport codes listed in Section \ref{sec:mccodes}.

This adoption of the object-oriented paradigm, carried out through well defined interfaces between
components and coupled with the open source character of the code, has allowed
scientists to extend the functionality of Geant4 in response to a wide variety
of experimental requirements.
Geant4 software design intrinsically supports multidisciplinarity through its
openness to extension and customization and is the key factor behind the wide
variety of research areas represented in the scientific publications using
Geant4, discussed in Section~\ref{sec:impact}.

The choice of C++ as the programming language for the implementation of Geant4
was motivated by objective considerations: its characteristic of being a
 de~facto standard, its use in industry, the availability of relevant development
tools and the expectation of a long lifetime.
These characteristics have contributed to the longevity of Geant4, avoiding the
risk of rapid obsolescence of the code and of its computational environment.

The use of mainframes meant that computing was conceived as a service provided
to a user sitting at a terminal  
and holding an account granted by a computing centre that had full control over
the programs available to the users.
Monte Carlo simulation fell into this scheme.
The availability of personal computers, which characterized the evolution of
high energy physics computing in the 1990s, 
enabled a scenario where the user became someone who worked and traveled with
her/his own computer, with the freedom of using the programs of her/his own
choice and the ability to connect with any other computer via the Internet.

RD44 effectively exploited the perspective of moving from computing based on a
mainframe architecture to personal computers and to heterogeneous High Performance Computing environments, from the TOP500 systems down to 
Beowulf  clusters, 
avoiding any technological lock-in as, for example, due to proprietary inter-node communication libraries.

The conception of Geant4 as a toolkit and its well-defined interfaces to external
systems for visualization, user interface, persistency etc., which shield users
from forced dependencies, grant users the autonomy to shape their simulation
environment as they deem appropriate to their experimental scenarios.
This resolution has also facilitated Geant4 users to devise high performance 
simulations based on commodity hardware for computationally intensive
experimental scenarios since the earliest Geant4 releases.
This approach is especially significant in the case of small research teams and
limited resources, which are largely represented among the publications
reporting scientific results based on Geant4 \cite{silver_siena23}.

The freedom of choice enabled by these strategic directions taken by RD44, which lets users
tailor the computational environment and the behaviour of the software,
including the ability to extend its functionality, is at the basis of the
outstanding amount and diversity of the research results associated with Geant4,
documented in Section~\ref{sec:impact}.

\subsection{Distributed Development}

The RD44 project introduced a novel approach to the development of a
general-purpose physics simulation tool as a geographically distributed project.

Previous simulation systems for particle transport were generally developed by
small teams, usually based at a research laboratory or at an academic institute:
notable examples are 
\cite{lanl_1974, 
calor_1977, 
egs3, 
geant_1978,
its_1986, 
ranft_1980,
fesefeldt_1985}.
The GEANT 3 \cite{geant321} code -- the predecessor of Geant4 at CERN -- was the
product of a CERN service unit operating in a dedicated computing department, 
with occasional contributions of a few visiting scientists.
High energy physics experiments in laboratories other than CERN commonly used
CERN service software, namely the CERN Program Library CERNLIB, which 
included the GEANT code.

The RD44 project was a a large international team of physicists, computer
scientists and engineers, based at several laboratories, research institutes and
universities worldwide.
The distributed development pattern was fundamental to involve scientists with
a wide spectrum of skills in all the physical, mathematical and
technological aspects of the complex problem domain, which contributed to both
the richness of Geant4 functionality and the quality of the toolkit.
The distributed character also introduced cooperative management of the
project, whose participants 
originated from different research areas and experiments that 
shared some common requirements, but differed in relation to
others. 
This trait was the seed of Geant4's multidisciplinarity and
significantly impacted the characteristics of the software.

Key features of RD44 distributed software development process
are documented in \cite{rd44-95, rd44-97, rd44-98}.
They included configuration management, which dealt with version control and
tags, incremental testing, which took into account package dependencies with
respect to the architectural design and related to traceability matrices, and
24-hour release cycles.
 
Distributed development was supported in RD44 by tools that represented the
state of the art at that time, such as CVS (Concurrent Version System) for version control and Rational Rose
for object-oriented analysis and design, along with tools developed within the
RD44 team in response to specific needs, such as unit testing and management of
24-hour release cycles. 
Their use evolved with time, along with general evolution in the field –
e.g., from CVS to SVN (Subversion) \cite{svn_book}, then to git
\cite{git_chacon} – and with the transition from RD44, a research project,
to the regime of the Geant4 Collaboration, characterized by maintenance and
support service to the user community.

\subsection{Software Development Process}
\label{sec:swprocess}

The research and development phase of Geant4 took four years (1995-1998),
following an iterative approach with progressive refinement of user
requirements, design, implementation and tests.
This software development strategy was critical to address the tremendous
complexity of the problem domain while performing a major transition to 
computational paradigms that were a novelty in the field of particle physics.

RD44's explicit adoption of a software development model represented an innovation by itself
in the particle physics environment.
The development process included several new activities -- for instance, a
24-hour testing and release cycle that today is common practice, but in the 1990's
it was mostly unknown in physics software and seldom adopted even in
professional software development environments.

Embracing an iterative and incremental software development process,
based on the Booch methodology \cite{booch_1990}, was the key to 
cope successfully with the aggressive development schedule imposed by the external
constraints of the LHC experimental program.
The RD44 project fulfilled all the milestones set by the CERN Scientific
Committees in the course of its lifetime and delivered the first public version
of Geant4 on time, with the required functionality.

At the end of RD44, the software development process conformed to the transition
from a fiercely innovative research pro\-ject to the activity oriented to
maintenance and user support in the environment of the Geant4 Collaboration.
Research on development methodologies is still pursued in some frontier
activities, whose innovative character benefits from the refinement of
development methods adopted in RD44 and the exploration of new methodologies,
with the support of pertinent tools. 
In this context, \cite{eclectic} reports the use of BPMN (Business Process Model and
Notation) \cite{bpmn_book} and of ArchiMate \cite{archimate32}, with 
focus on identifying the role of these technologies in physics research.

\subsection{Usability and sustainability}
\label{sec:usability}

The adoption of new computing paradigms and 
the use of an unfamiliar programming language required clever
efforts to ensure the usability of Geant4 in experimental physics applications.
Several actions, planned in the course of the RD44 project and carried out since
the initial Geant4 release, have contributed to Geant4 usability: the inclusion in the
toolkit of a wide collection of application examples, encompassing the
illustration of various features as well as simulations of realistic
experimental scenarios, a series of introductory seminars at laboratories and
universities, and several presentations at international scientific conferences
and local meetings to inform the diverse user communities about novelties and
achievements especially relevant to their field.
Today, learning Geant4 is included in the academic programs of physics and
engineering degrees at many universities worldwide.

Under the regime of maintenance of the international Geant4
collaboration, the code has been updated to align with the C++11 and C++17
standards, to comply with the evolution of compilers and to exploit the
multithreading capabilities available in recent hardware \cite{g4nim2}.
Thanks to the sound software design established in the RD44 project, the
original functionality of Geant4 has been extended in many areas with the
contribution of both formal collaboration members and users.
Source code and application examples donated by users are maintained by 
collaboration members to ensure their long term availability when long-term
commitment of the original contributors is not possible.

\section{Conclusion and Future Perspectives}
\label{sec:conclusion}

Geant4 is still actively used 25 years since its initial release. 
As shown in Section \ref{sec:impact}, the number of citations of its main
reference paper and of patents based on it has been steadily growing.

Based on its usage and citations, Geant4 has been the most impactful Monte Carlo
software code in an extended scientific community for 25 years.
It has been applied in a wide variety of scientific fields, ranging from high
energy physics to nuclear medicine to space science to archaeology.
Some key factors have contributed to its longevity and broad applicability:
the code has been open source, allowing users to contribute improvements and
extensions of functionality;
it used an object-oriented software paradigm, which provided a toolkit with
large flexibility in applying it to a variety of experimental scenarios;
it was written in a modern, widely used, long-lived programming language;
the development team was broad-based, both geographically and scientifically.
The directions taken by RD44, which today represent good software development practice,
represented pioneering principles and cutting-edge technology 
the 1990s in the field of high energy physics.

Geant4 is expected to remain a reference simulation software into the future,
thanks to its widespread use in many application areas, its openness to
improvements and extension in response to
experimental requirements, and the Geant4 collaboration's continual commitment
to its maintenance.
An extensive discussion of Geant4 future perspectives exceeds the scope of this
paper; below are a few topics of current interest in the domain of Monte Carlo
particle transport, which are likely to be the object of research in the
field in the coming years.

The need to increase simulation throughput is common to various 
application environments, from experiments at high luminosity colliders to
personalized medical treatments.
Ongoing work to introduce sub-event parallelism \cite{geant4_112} and
to demonstrate the feasibility and efficiency of adopting the GPU computing paradigm
involves revisiting computationally intensive parts of Geant4 code.
Increasing heterogeneity in computing architectures, where a single CPU can
provide 128 cores (e.g. AMD EPYC Bergamo) or more by merging CPU and GPU cores
(e.g. AMD Instinct MI300A), would demand further investment to keep up with the
evolution of computing hardware.

State-of-the-art computational performance is needed not only by large scale
Geant4-based simulations, but also for Uncertainty Quantification, an
emerging discipline in predictive computational science \cite{mcclarren_2018}.
In the domain of particle transport, it represents the ability to
objectively quantify the degree of reliability of the outcome of a simulation on
the basis of the uncertainties in the simulation model \cite{kennedy_2001} and
involves developments for the propagation of errors from the data used by
the simulation to its results.
This capability is especially relevant to critical applications, e.g. in
medicine and radiation protection, and to scenarios where either cost or
practical constraints prevent assessing the reliability of complex simulated
observables by means of direct experimental measurements.
Multidisciplinary research, involving mathematical, statistical, computational
and epistemological aspects, is required to this end.

Uncertainty Quantification is closely related to the validation of Geant4 physics
``ingredients'': atomic and nuclear parameters, interaction
cross sections, modelling methods and algorithms, etc.
Currently, only a small fraction of their validation tests use rigorous
statistical methods, able to produce objective, quantitative results.
Experimental measurements needed for validation are scarce or
insufficiently precise in some areas; the new paradigm of computational
simulation promoted in \cite{post_2005}, which argued the need of funding
experiments explicitly conducted for code validation, is far from established.
Progress in this domain would be beneficial.

Two key factors characterized the creation of Geant4: the vision that shaped the
RD44 project as original, innovative scientific research rather than as a mere
computational service, and the talent to make the most of the skills and
diversity of all participants towards a shared goal.
They were crucial to cope successfully with the technical and environmental
challenges outlined in Section \ref{sec:sweng}.
They would still be valuable guidelines for the future of Geant4.

\section*{Acknowledgements} 
The authors acknowledge many valuable discussions with RD44 members and Geant4 users.
The CERN Library has been especially helpful to retrieve hard-to-find literature.

\appendix
\section{RD44 Project}
\label{app:rd44}
The link to the list of RD44 members in the final project report \cite{rd44-98}
is broken; the list is below to properly credit the RD44 members for their
contribution to the development of Geant4.

The members of the RD44 project were: 
J. Allison (Manchester Univ.), 
K. Amako (KEK), 
N. Ameline (JINR), 
J. Apostolakis (CERN), 
C. Arnault (LAL, Orsay), 
P. Arce (Univ. Santander), 
M. Asai (Hiroshima Inst.  of Tech.), 
D. Axen (UBC), 
G. Ballocchi (INFN Padova), 
G. Barrand (LAL, Orsay)
% ...

 \bibliographystyle{elsarticle-num} 
 \bibliography{bibliography}





\end{document}